\documentclass[12pt,epsf]{article}
\usepackage{graphicx, amsmath}

\begin{document}

\sloppy
\begin{flushright}{SIT-HEP/TM-45}
\end{flushright}
\vskip 1.5 truecm
\centerline{\large{\bf Modulated Inflation (@SUSY08)}}
\vskip .75 truecm
\centerline{\bf Tomohiro Matsuda\footnote{matsuda@sit.ac.jp}}
\vskip .4 truecm
\centerline {\it Laboratory of Physics, Saitama Institute of Technology,}
\centerline {\it Fusaiji, Okabe-machi, Saitama 369-0293, 
Japan}
\vskip 1. truecm
\makeatletter
\@addtoreset{equation}{section}
\def\theequation{\thesection.\arabic{equation}}
\makeatother
\vskip 1. truecm

\begin{abstract}
\hspace*{\parindent}
We consider cosmological perturbations caused by modulated
inflaton velocity. 
During inflation, the inflaton motion is damped and the velocity is
determined by the parameters such as couplings or
masses that may depend on light fields(moduli).
The number of e-foldings is different in different patches if there are
spatial fluctuations of such parameters.
Based on this simple idea, we consider ``modulated inflation''
in which the curvature perturbation is generated by the fluctuation of
the inflaton velocity. This talk is based on our recent papers\cite{Matsuda:Modulatedinflation3}. 
\end{abstract}
\newpage

\section{Introduction}

\begin{figure}[ht]
 \begin{center}
\begin{picture}(400,250)(0,0)
\resizebox{15cm}{!}{\includegraphics{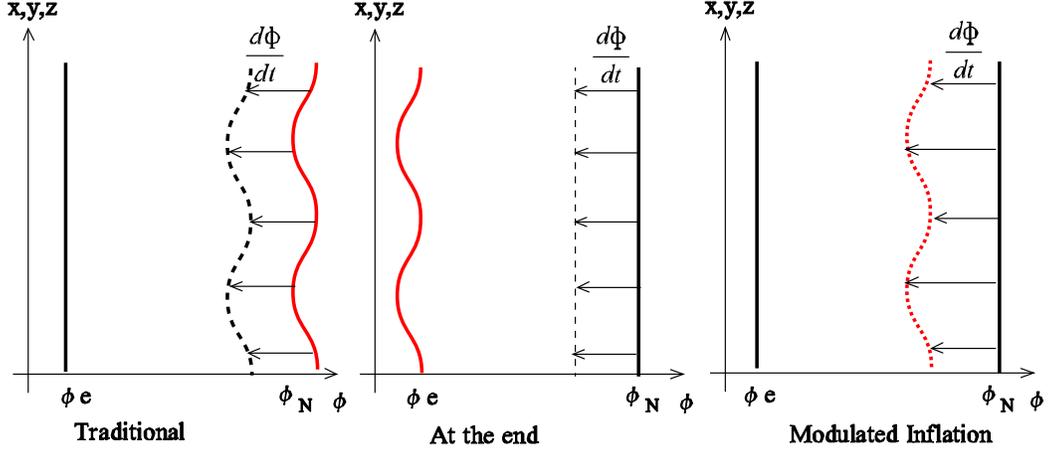}} 
\end{picture}
 \caption{In the first(Traditional) and the second(At the end) scenarios, $\delta t$ is caused by
 the fluctuation of the distance between the start-line at $\phi_N$ and
 the goal-line at $\phi_e$, while in the last scenario $\delta t$ is
 caused by $\delta \dot{\phi}$.}
\label{fig:normal}
 \end{center}
\end{figure}

Let us first discuss common ideas for cosmological perturbations
in terms of the $\delta N$ formalism.
Assuming that $H$ is a constant during inflation, the formula for
$\delta N$ is given by $\delta N \simeq H \delta t$, where $\delta t$ is
the fluctuation related to the time passed after horizon crossing.
According to the traditional inflationary scenario, the curvature
perturbation is generated at the horizon crossing\footnote{See the first
picture in Fig.\ref{fig:normal}.}. 
Another idea for the cosmological perturbation is
``at the end'' scenario\cite{Attheend}\footnote{See the second picture in
Fig.\ref{fig:normal}.}.
In the `` at the end'' scenario, the curvature perturbation is generated
at the end, where the fluctuation of the goal-line is induced by
the light field (${\cal M}$); 
$\delta \phi_e \simeq \phi_e' \delta {\cal
M}$. 
Of course, considering the fluctuation of the
distance ($\delta\phi$) to obtain $\delta t$ is very natural, as it
obviously causes the fluctuation of the number of e-foldings $\delta N$.
However, we know that the time elapsed after horizon crossing
$\sim |t_N-t_e|$ depends not only on the distance
$\sim|\phi_N-\phi_e|$ but also on the inflaton velocity $\dot{\phi}$. 
Therefore, if the inflaton velocity depends on a light field, 
the fluctuation of the light field may lead to
 the fluctuation of the inflaton velocity; $\delta\dot{\phi}\sim
 (\dot{\phi})'\delta {\cal M}$, which eventually causes $\delta t$ and
 $\delta N$. 
The effect would be obvious when
(1) many massless degrees of freedom appear in the inflaton
trajectory, or
(2) the inflaton mass is slightly larger than the Hubble parameter and
there is no significant perturbation from $\delta \phi_N$.
Based on the above simple speculation, we will focus on
modulated\cite{Kofman:modulated} inflaton velocity, 
where spatial fluctuation of the time elapsed during inflation
$\delta t$ is caused by the fluctuation of the 
inflaton velocity.\footnote{See the red-dotted line in the last picture
in Fig.\ref{fig:normal}.} 
Moreover, if the fundamental parameters are determined by the moduli in
the underlying (string) theory, it would be natural to think that the
Planck scale may also depend on moduli.
What happens if the Planck scale is determined by moduli fields that may
have fluctuations during inflation?
We will see that the perturbation related to $\delta M_p(\cal M)$ is 
explained in terms of $\delta \dot{\phi}$.
Let us see more details of this ``modulated velocity scenario'',
considering simple examples.

\subsection{Modulated velocity from the inflaton potential}

For the first example, we consider standard kinetic terms and 
conventional interaction term in the potential, $V(\phi, {\cal M})$.
In this case, the definition of the number of e-foldings for constant
$H$ is 
\begin{equation}
N = \int H dt =\int H\frac{\dot{\phi}d\phi +\dot{\cal M}
d{\cal M}}{\dot{\phi}^2 + \dot{\cal M}^2}.
\end{equation}
If there is no bend in the trajectory\cite{multi-inflation}, the
perturbation related to the inflaton velocity is expanded as   
\begin{equation}
\label{delta-vel}
\delta N 
\simeq -\int^{\phi_N}_{\phi_e}\frac{H}{\dot{\phi}^2}\left(
\delta\dot{\phi}-\dot{\phi}A\right)d\phi,
\end{equation}
where we consider linear scalar perturbations of the metric,
$ds^2 = -(1+2 A)dt^2 + 2aB_idx^idt + a^2\left[(1-2\psi)\delta_{ij}+2E_{ij}
\right]dx^i dx^j$.
From the conventional energy and momentum constraints we can find
\begin{equation}
\dot{\phi}\left( \dot{\delta \phi}-\dot{\phi}A\right)
-\ddot{\phi}\delta\phi
\simeq \dot{\phi}\left(\dot{\delta \phi}-\dot{\phi}A\right) 
\propto \frac{k^2}{a^2},
\end{equation}
which suggests that the factor $\dot{\delta \phi}-\dot{\phi}A$ appearing
in the delta-N formula decays after horizon crossing.
Therefore, the factor $e^{-2Ht}$ must be included in the
calculation, even if the perturbation $\dot{\phi}\delta\dot{
\phi}$ itself is supposed to be a constant.
Of course, $e^{-2Ht}$ in the integral does not lead to
the exponential suppression after the integration. It is easy to
see that there is no exponential suppression for large $N$($\sim Ht$),
\begin{equation}
\int^{t_e}_0 \delta C  H e^{-2Ht} dt \simeq \frac{1}{2}
\delta C,
\end{equation}
where the actual suppression factor is not significant.
In fact, the corrections from terms proportional to $k^2/a^2$ have
been disregarded in previous studies, since $k^2/a^2$ is obviously small
at a distance.  
However, if they appear in the equation of $\dot{\cal R}$, 
these terms may yield significant correction to ${\cal R}$
 after integration, as we can see easily from the above equation.

\subsection{Modulated velocity from the inflaton kinetic term}

Next, we consider moduli-dependent kinetic term $\sim \frac{1}{2}
\omega({\cal M}) g^{ab}\phi_{,a}\phi_{,b}$.
In this case, the number of e-foldings is
\begin{equation}
N = \int H dt \simeq \int \frac{H}{\dot{\phi}}d\phi\simeq
\int \frac{3H^2}{V_\phi} \omega d\phi.
\end{equation}
Again, we assume no bend.
$\delta N$ that is caused by the moduli perturbation is 
\begin{equation}
\delta N \simeq \int \frac{3H^2}{V_\phi}
\omega' \delta {\cal M} d\phi\simeq 
 \frac{\omega' N}{\omega}  \delta {\cal M}.
\end{equation}
Note that unlike the perturbation caused by the
 potential, the constraints from the energy and momentum 
does not yield $\frac{k^2}{a^2}$ factor for the perturbation
 related to the kinetic term\cite{kinetic-term}.

\section{Pure case}

Let us see what happens if both boundaries are completely flat
 while there is modulated velocity. 
Here we consider fast-roll inflation with hybrid-type
 potential\cite{Matsuda:eta-brane, Matsuda:nontach}. 
Inflaton may have large mass ($m_\phi \simeq O(1)H$) due to the
$\eta$-problem. 
Then $\delta \phi$ decays or cannot cross
the horizon during inflation. 
Even in this case, non-oscillationary (fast-roll) inflation is
 possible if the friction is significant.
We consider the hybrid-type potential
\begin{equation}
V(\phi,\sigma)=\lambda\left(\sigma^2-v^2\right)^2
+\frac{1}{2}g^2\phi^2\sigma^2 + V(\phi),
\end{equation}
where $\phi$ is the inflaton and $\sigma$ is the trigger field.
Here the end of inflation expansion occurs at
\begin{equation}
\phi_e=\frac{\sqrt{\lambda}v}{g},
\end{equation}
and the number of e-foldings is given by
\begin{equation}
N=\frac{1}{F_\phi}\log\frac{\phi_N}{\phi_e},
\end{equation}
where $F\equiv \frac{3}{2}\left(1-\sqrt{1-4m^2/9H^2}\right)$. 
Considering the factor $\sim k^2/a^2$, we find for $m\simeq H$,
\begin{equation}
\delta N_{(\cal M)} \sim m' \delta {\cal M},
\end{equation}
where $m'$ is the derivative of $m$ with respect to ${\cal M}$.
More specific result is obtained for  
\begin{equation}
m^2({\cal M})\equiv m^2_0\left[1+\beta\log({\cal M}/M_*)\right],
\end{equation}
where $\delta N$ is given by
\begin{equation}
\delta N_{(\cal M)} \simeq \beta \left(\frac {\delta{\cal M}}{{\cal M}}\right). 
\end{equation}
Since the mass of ${\cal M}$ is smaller than
$H_I$ during inflation, we find the condition
\begin{equation}
\label{another-slow}
m^2_{\cal M}\simeq \beta m^2_0\left(\frac{\phi_N}{\cal M}\right)^2
< H_I^2.
\end{equation}
In this case, the non-Gaussianity parameter is
\begin{equation}
f_{nl}=-\frac{5}{6}\frac{N''}{(N')^2}\propto
\frac{1}{ \beta},
\end{equation}
which can be large and may take either
sign.\footnote{Curvaton and inhomogeneous
preheating may lead to
curvature perturbations with significant
non-Gaussianity\cite{Matsuda:curvatons, 
Matsuda:Inhomogeneouspreheating}.}

\section{Adding significant non-gaussianity}

Our question is 
 ``Is it possible to add significant 
non-gaussianity to the standard inflationary perturbation after horizon
crossing?'' 
Our answer is ``Yes''.
To show how to add non-gaussianity to the conventional perturbation, 
we consider hybrid inflation with standard $\sim g^2 \phi^2{\cal M}^2$
 interaction.  
During inflation, hybrid inflation has the effective potential
\begin{equation}
V(\phi,{\cal M})=V_0 + \frac{1}{2}m_\phi^2 \phi^2
+g^2(\phi-\phi_{ESP})^2{\cal M}_i^2,
\end{equation}
where ${\cal M}_i$ become massless near the enhanced symmetric point(ESP)
 at $\phi_{ESP}$. 
The perturbation of the inflaton velocity caused by the number of n
massless excitation is 
\begin{eqnarray}
\delta\dot{\phi} &\simeq& \frac{2ng^2 (\phi-\phi_{ESP})
(\delta {\cal M})^2}{3H} \end{eqnarray}
where the first order perturbation vanishes.
The second order perturbation at the ESP adds significant
non-gaussianity to the perturbation;
\begin{equation}
\hat{f}_{NL} \simeq \frac{\delta{\dot{\phi}}/\dot{\phi}}
{(H^2/\dot{\phi})^2}\simeq 
\frac{n\eta_\phi\eta_{\cal M}\phi}{\phi-\phi_{ESP}}.
\end{equation}
Since the non-gaussianity is uncorrelated,
the usual non-linear parameter $f_{NL}$ is given by $f_{NL}\sim
(\hat{f}_{NL}/1300)^3$\cite{uncorrelated-Lyth}. 
We conclude that we can add significant
non-gaussianity to the standard perturbation after horizon crossing.
\begin{figure}[ht]
 \begin{center}
\begin{picture}(400,250)(0,0)
\resizebox{13cm}{!}{\includegraphics{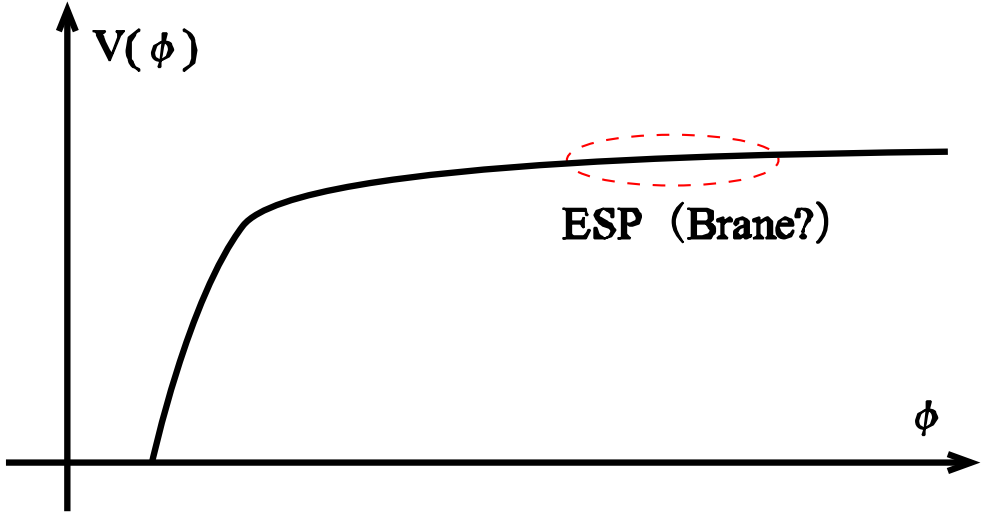}} 
\end{picture}
\caption{ESP may appear in the inflaton trajectory\cite{beauty}.}
\label{fig:esp}
 \end{center}
\end{figure}

\section{Modulated Planck scale}

Finally, we briefly mention the consequence of modulated Planck scale in
terms of the delta-N formalism.
We consider a light scalar
field $\hat{\cal M}$ coupled to gravity. 
The model is given by the action
\begin{eqnarray}
S&=&\int d^4 x \sqrt{-g} \left[
f(\hat{\cal M}) R -g(\hat{\cal M}) \left(\nabla \hat{\cal M}\right)^2
\right.\nonumber\\
&&\left.-\frac{1}{2} \left(\nabla \phi\right)^2 -V(\phi)\right].
\end{eqnarray}
Considering (for simplicity) Jordan-Brans-Dicke theory,
the action in the Einstein frame has the
moduli-dependent kinetic term $\sim \frac{1}{2} \omega({\cal M}) 
g^{ab}\phi_{,a}\phi_{,b}$, where
\begin{equation}
\omega({\cal M})
=\exp\left(-\beta\kappa {\cal M}\right)
\end{equation}
and the potential
\begin{equation}
V=\omega^2 W(\phi).
\end{equation}
Note that there are both sources for the velocity perturbation, from the
potential  
and the kinetic term.
The perturbation caused by the potential has the factor $k^2/a^2$,
while the one from the kinetic term does not.

\section{Summary}

We considered cosmological perturbations caused by modulated inflaton
velocity. 
The velocity perturbation has been disregarded in previous studies for
multi-field inflation.
However, the perturbation may lead to significant results, as we
have shown in this presentation.
The sources of such perturbations are clear in the $\delta N$ formalism.
Important results are:
\begin{itemize}
\item Small deviation from the standard perturbation may be explained by
      the modulated velocity.
\item The curvature perturbation can be generated after horizon crossing
even if the both boundaries are completely flat.
\item It is possible to add significant non-gaussianity to the
       conventional perturbations after horizon crossing.
\end{itemize}

\section{Acknowledgment}
We wish to thank the SUSY'08 staff and
participants for their kind hospitality.


\begin{thebibliography}{9}
\bibitem{Matsuda:Modulatedinflation3}
  T.~Matsuda,
  ``Modulated Inflation,''
  [arXiv:0801.2648];
  T.~Matsuda,
  ``Running spectral index from shooting-star moduli,''
  JHEP {\bf 0802}, 099 (2008)
  [arXiv:0802.3573].
\bibitem{Attheend}
  F.~Bernardeau, L.~Kofman and J.~P.~Uzan,
  ``Modulated fluctuations from hybrid inflation,''
  Phys.\ Rev.\  D {\bf 70}, 083004 (2004)
  [arXiv:astro-ph/0403315];
  D.~H.~Lyth,
  ``Generating the curvature perturbation at the end of inflation,''
  JCAP {\bf 0511}, 006 (2005)
  [arXiv:astro-ph/0510443].
\bibitem{Kofman:modulated}
  L.~Kofman,
  ``Probing string theory with modulated cosmological fluctuations,''
  arXiv:astro-ph/0303614.
\bibitem{multi-inflation}
  F.~Bernardeau and J.~P.~Uzan,
  ``Non-Gaussianity in multi-field inflation,''
  Phys.\ Rev.\  D {\bf 66}, 103506 (2002)
  [hep-ph/0207295].
\bibitem{kinetic-term}
  T.~Matsuda,
  ``Modulated inflation from kinetic term,''
  JCAP {\bf 0805}, 022 (2008)
  [arXiv:0804.3268];
  J.~Garcia-Bellido and D.~Wands,
  ``Constraints from inflation on scalar - tensor gravity theories,''
  Phys.\ Rev.\  D {\bf 52}, 6739 (1995)
  [gr-qc/9506050].
\bibitem{Matsuda:eta-brane}
  T.~Matsuda,
  ``Brane inflation without slow-roll,''
  JHEP {\bf 0703}, 096 (2007)
  [astro-ph/0610402];
  T.~Matsuda,
  ``Elliptic inflation: Generating the curvature perturbation without
  slow-roll,''
  JCAP {\bf 0609}, 003 (2006)
  [hep-ph/0606137];
  T.~Matsuda,
  ``Thermal hybrid inflation in brane world,''
  Phys.\ Rev.\  D {\bf 68}, 047702 (2003)
  [hep-ph/0302253];
  T.~Matsuda,
  ``Topological hybrid inflation in brane world,''
  JCAP {\bf 0306}, 007 (2003)
  [hep-ph/0302204].
\bibitem{Matsuda:nontach}
  T.~Matsuda,
  ``Non-tachyonic brane inflation,''
  Phys.\ Rev.\  D {\bf 67}, 083519 (2003)
  [hep-ph/0302035];
  T.~Matsuda,
  ``F-term, D-term and hybrid brane inflation,''
  JCAP {\bf 0311}, 003 (2003)
  [hep-ph/0302078].
\bibitem{Matsuda:curvatons}
  T.~Matsuda,
  ``Curvaton paradigm can accommodate multiple low inflation scales,''
  Class.\ Quant.\ Grav.\  {\bf 21}, L11 (2004)
  [hep-ph/0312058];
  T.~Matsuda,
  ``Hilltop Curvatons,''
  Phys.\ Lett.\  B {\bf 659}, 783 (2008)
  [arXiv:0712.2103];
  T.~Matsuda,
  ``Hybrid Curvatons from Broken Symmetry,''
  JHEP {\bf 0709}, 027 (2007)
  [arXiv:0708.4098].
  T.~Matsuda,
  ``Topological curvatons,''
  Phys.\ Rev.\  D {\bf 72}, 123508 (2005)
  [hep-ph/0509063].
\bibitem{Matsuda:Inhomogeneouspreheating}
  T.~Matsuda,
  ``Generating the curvature perturbation with instant preheating,''
  JCAP {\bf 0703}, 003 (2007)
  [hep-th/0610232];
  T.~Matsuda,
  ``Generating curvature perturbations with MSSM flat directions,''
  JCAP {\bf 0706}, 029 (2007)
  [hep-ph/0701024];
  T.~Matsuda,
  ``Cosmological perturbations from inhomogeneous preheating and
  	multi-field trapping,''
  JHEP {\bf 0707}, 035 (2007)
  [arXiv:0707.0543].
\bibitem{uncorrelated-Lyth}
L.~Boubekeur and D.~H.~Lyth,
  ``Detecting a small perturbation through its non-Gaussianity,''
  Phys.\ Rev.\  D {\bf 73}, 021301 (2006)
  [astro-ph/0504046].
\bibitem{beauty}
  L.~Kofman, A.~Linde, X.~Liu, A.~Maloney, L.~McAllister and
	E.~Silverstein, 
  ``Beauty is attractive: Moduli trapping at enhanced symmetry points,''
  JHEP {\bf 0405}, 030 (2004)
  [hep-th/0403001].
\end{thebibliography}
\end{document}